\documentclass[preprint,prl,showpacs,showkeys,floatfix]{revtex4}
\newcommand{\bfm}[1]{\mbox{\boldmath${#1}$}}
\usepackage{graphicx}
\begin{document}
\title{Generalized Kinetic Theory of Electrons
and Phonons:\\ Models, Equilibrium, Stability}
\author{A. Rossani\footnote{Corresponding author}}\email{alberto.rossani@polito.it}
\author{A.M. Scarfone}\email{antonio.scarfone@polito.it}
\affiliation{Istituto Nazionale di Fisica della Materia and
\\Dipartimento di Fisica -
 Politecnico di Torino\\ Corso Duca degli Abruzzi 24, 10129 Torino, Italy.}
\date{\today}

\begin {abstract}
In the present paper our aim is to introduce some models for the
generalization of the kinetic theory of electrons and phonons
(KTEP), as well as to study equilibrium solutions and their
stability for the generalized KTEP (GKTEP) equations. We consider
a couple of models, relevant to non standard quantum statistics,
which give rise to inverse power law decays of the distribution
function with respect to energy. In the case of electrons in a
phonon background, equilibrium and stability are investigated by
means of Lyapounov theory. Connections with thermodynamics are
pointed out.
\end{abstract}
\pacs{05.20.Dd; 05.60.Cd; 72.10.-d} \keywords{Kinetic theory;
Electrons and phonons}\maketitle

\section{Introduction}

As pointed out by Koponen, fractal or inverse power law
distributions are of interest in modeling various meaningful
situations in solid state physics \cite{1}. An example, treated in
ref. \cite{1}, is the thermalization of a nonequilibrium
electron-phonon system.\\
Until recently, however, there has been little guidance on how to
generalize the kinetic theory of electrons and phonons obeying
non-Gibbsian statistics. In a very recent paper \cite{2} a
generalized kinetic theory for electrons and phonons has been
proposed. Modified collision terms were introduced, in order to
allow applications not only to electrons, but also to other
particles (obeying a
general statistics) which interact with a crystal lattice.\\
In the present paper our aim is to introduce some models for the
generalization of the KTEP equations as well as to study
equilibrium and stability for the GKTEP equations. We introduce a
couple of models for the GKTEP, relevant to non standard quantum
statistics recently proposed in the literature, and compare them
from the point of view of Kinetic Theory. Both of them give rise
to inverse power decays of the distribution function with respect
to energy. In absence of external disturbances, it is possible to
treat phonons as background at equilibrium. This assumption
implies a drastic simplification of the GKTEP, since we have just
to deal with the equation for electrons. Equilibrium solutions
and their stability are investigated by means of Lyapunov's
theory. The connection between mathematical results and
Thermodynamics is shown.

First we recall the GKTEP equations and their stationary
solutions. Let $N_g=N_g(\bfm k,\,{\bfm x},\,t)$ be the
distribution function of phonons (quasi-momentum  $\bfm k$,
energy $\omega_g(\bfm k)$ of branch $g$ of the phonon spectrum and
$n_{\bf p}=n({\bfm p},\,{\bfm x},\,t)$ the distribution function
of electrons
[quasi-momentum ${\bfm p}$, energy $\epsilon({\bfm p})$].\\
The generalized kinetic equations for phonons and electrons read
\cite{2}:
\begin{eqnarray}
&& \frac{\partial\,N_g}{\partial\,
t}+{\bfm u}_g\cdot\frac{\partial\,N_g}{\partial\,{\bfm x}}=
\left(\frac{\partial\,N_g}{\partial\,
t}\right)_{pp}+\left(\frac{\partial\,N_g}{
\partial\,t}\right)_{pe} \ ,\\
&&\frac{\partial\,n_{\bf p}}{\partial\, t}+{\bfm
v}\cdot\frac{\partial\,n_{\bf p}}{\partial\,{\bfm x}} -e\,{\bfm
E}\cdot\frac{\partial\,n_{\bf p}}{\partial\,{\bfm
p}}=\left(\frac{\partial\,n_{\bf p}}{\partial\, t}\right)_{ep} \ ,
\end{eqnarray}
where $ {\bfm u}_g=\partial\,\omega_g/\partial\,\bfm k$ and
${\bfm v}=\partial\,\epsilon_{\bf p}/\partial\,{\bfm p}$. The right hand
sides are given by
\begin{eqnarray}
\nonumber \left(\frac{\partial\,N_g}{\partial\,
t}\right)_{pp}&=&-\,\int\Big\lbrace{1\over2}\,
\sum_{g_1g_2}w_{pp}({\bfm k}_1,\,{\bfm k}_2\rightarrow{\bfm k})\,\delta(\omega_g-\omega_{g_1}-\omega_{g_2})\\
\nonumber &&\times[\Phi(N_g)\,\Psi(N_{g_1})\,\Psi(N_{g_2})-
\Phi(N_{g_1})\,\Phi(N_{g_2})\,\Psi(N_g)]\\
\nonumber &&
+\sum_{g_1g_3}w_{pp}({\bfm k},\,{\bfm k}_1\rightarrow{\bfm k}_3)\,\delta(\omega_{g_3}-\omega_g-\omega_{g_1})\\
&&\times[\Phi(N_{g_3})\,\Psi(N_g)\,\Psi(N_{g_1})-
\Phi(N_g)\,\Phi(N_{g_1})\,\Psi(N_{g_3})]\Big\rbrace\,\frac{d{\bfm k}_1}{
8\,\pi^3} \ ,\\
\nonumber &&\\ \nonumber \left(\frac{\partial N_g}{\partial\,
t}\right)_{pe}&=&2\int
w_{pe}({\bf p}\rightarrow{\bf p}^\prime,\,{\bfm k})\,\delta(\epsilon_{\bf p^\prime}+\omega_g-\epsilon_{\bf p})\\
&&\times[\varphi(n_{\bf p})\,\psi(n_{\bf
p^\prime})\,\Psi(N_g)-\Phi(N_g)\,\varphi(n_{\bf p^\prime})\,
\psi(n_{\bf p})]\,\frac{d{\bfm p}}{8\,\pi^3} \ ,\\
\nonumber &&\\
\nonumber \left(\frac{\partial n_{\bf p}}{\partial t}\right)_{ep}
&=&\sum_g\int\Big\lbrace
w_{ep}({\bfm p}^\prime,\,{\bfm k}\rightarrow{\bfm p})\,\delta(\epsilon_{\bf p}-\epsilon_{\bf p^\prime}-\omega_g)\\
\nonumber &&\times[\Phi(N_g)\,\varphi(n_{\bf p^\prime})\,\psi(n_{\bf
p})
-\varphi(n_{\bf p})\,\psi(n_{\bf p^\prime})\,\Psi(N_g)]\\
\nonumber &&
+w_{ep}({\bfm p}^\prime\rightarrow{\bfm p},\,{\bfm k})\,\delta(\epsilon_{\bf p}+\omega_g-\epsilon_{\bf p^\prime})\\
&&\times[\varphi(n_{\bf p^\prime})\,\psi(n_{\bf p})\,\Psi(N_g)
-\Phi(N_g)\,\varphi(n_{\bf p})\,\psi(n_{\bf
p^\prime})]\Big\rbrace\,\frac{d{\bfm k}} {8\,\pi^3} \ ,\label{15}
\end{eqnarray}
where the $w$'s are transition probabilities and
electron-electron interactions have been neglected. Observe that
in the limit
\begin{eqnarray}
&&\Psi(N_g) \ \to\ 1+N_g \ ,\hspace{2cm} \Phi(N_g)\ \to\ N_g \ ,\\
&&\psi(n_{\bf p})\ \to\ 1-n_{\bf p} \ ,\hspace{2.1cm} \varphi(n_{\bf p})\ \to\ n_{\bf p} \ ,
\end{eqnarray}
the Bloch-Boltzmann-Peierls (BBP) equations \cite{2} are
recovered. The non negative functions $\Psi$, $\Phi$ for phonons
[$N_g\in[0,\,\infty)]$, and $\psi$, $\varphi$ for electrons
($n_{\bf p}\in[0,\,1]$) represent the arrival and
departure state availability, respectively. Moreover we assume:\\
{\it i}) $\Phi/\Psi$ and $\varphi/\psi$ are monotonically
increasing functions of $N_g$ and $n_{\bf p}$, respectively \cite{2};\\
{\it ii}) the following conditions hold
\begin{eqnarray}
&&\varphi(0)=0 \ ,\ \ \Phi(0)=0 \ ,\label{c1}\\
&&\psi(0)=1 \ ,\ \ \Psi(0)=1 \ ,\label{c2}\\
&&\psi(1)=0 \ ,\ \ \Psi(+\infty)=+\infty\label{c3} \ .
\end{eqnarray}
Conditions (\ref{c1}) mean that no transition occurs when the
initial state is empty, conditions (\ref{c2}) mean that the
emptiness of the arrival state neither inhibits nor enhances a
transition and conditions (\ref{c3}) are characteristic of
fermions and bosons, respectively.\\ In \cite{2} it has been shown
that, implicitly, the equilibrium solutions of the generalized
system are:
\begin{eqnarray}
&&\ln\frac{\Phi(N_g^\ast)}{\Psi(N_g^\ast)}=-\frac{\omega_g}{T} \ ,\label{1}\\
&&\ln\frac{\varphi(n_{\bf p}^\ast)}{\psi(n_{\bf p}^\ast)}=-\frac{\epsilon_{\bf
p}-\mu}{T} \ ,\label{2}
\end{eqnarray}
($\ast$ means "at equilibrium"), where $T$ is the absolute
temperature of the mixture and $\mu$ is the chemical potential of
the electron gas.

\section{Models for the generalization} Preliminarily, consider
certain particles, whose equilibrium distribution function is
$f$, which obey a statistics described by the parameter
$\lambda\in[-1,+1]$. Hereinafter hatted quantities are referred
to these particles. In particular, for $\lambda=\ -1,\ 0,\ +1,$
we have fermions, classical particles, and bosons, respectively.
Suppose, we invert this distribution function with respect to
$\exp(-{\cal E})$ [where ${\cal E}=(\hat\epsilon-\hat\mu)/T$]
than we obtain:
\begin{eqnarray}
\exp(-{\cal E})=F(f,\,\lambda) \ .\label{3}
\end{eqnarray}
In particular, for electrons ${\cal E}=(\epsilon_{\bf p}-\mu)/T$, for
phonons ${\cal E}=\omega_g/T$, since, in this case, $\hat\mu=0.$ The
statistics of the particles we deal with is described also by the
couple of function $\hat\varphi(f,\,\lambda)$ and
$\hat\psi(f,\,\lambda)$, essentially positive, and obeying the
following conditions:
\begin{eqnarray}
\hat\varphi(0,\,\lambda)=0 \ ,\hspace{2.3cm} \hat\psi(0,\,\lambda)=1 \
.
\end{eqnarray}
In particular we have
\begin{eqnarray}
&&\varphi(n_{\bf p})=\hat\varphi(n_{\bf p},\,-1) \ ,\hspace{1.1cm}
\psi(n_{\bf p})=\hat\psi(n_{\bf p},\,-1) \
,\\
&&\Phi(N_g)=\hat\varphi(N_g,\,+1) \ ,\hspace{1cm}
\Psi(N_g)=\hat\psi(N_g,\,+1) \ .
\end{eqnarray}
From (\ref{1}) and (\ref{2}) we know that at equilibrium
\begin{eqnarray}
\frac{\hat\varphi}{\hat\psi}=\exp(-{\cal E}) \ .\label{4}
\end{eqnarray}
Now, Eqs. (\ref{3}) and (\ref{4}) give
\begin{eqnarray}
\frac{\hat\varphi}{\hat\psi}=F(f,\,\lambda) \ .\label{5}
\end{eqnarray}
Let us introduce the following assumptions:\\
{\it i}) $\hat\varphi/\hat\psi$ is given by Eq. (\ref{5}) not only at
equilibrium;\\
{\it ii}) $\hat\varphi(f,\,\lambda)$ does not depend on $\lambda$,
and we
take $\hat\varphi(f,\,\lambda)=\hat\varphi(f,\,0)$;\\
{\it iii}) $\hat\psi(f,\,0)=1$.\\
Properties {\it i}), {\it ii}) and {\it iii}) actually hold for
standard statistics (Fermi-Dirac, Maxwell-Boltzmann, and
Bose-Einstein) and, more in general, we postulate them. From Eq.
(\ref{5}), by accounting for {\it ii}) and {\it iii}), we get
\begin{eqnarray}
\hat\varphi=F(f,\,0) \ ,\hspace{2cm} \hat\psi=\frac{F(f,\,0)}{
F(f,\lambda)} \ .\label{6}
\end{eqnarray}

We introduce now two models which account for an inverse power
law decay of the distribution function with respect to energy:\\

{\bf 1}) The first model that we consider has been proposed by
B\"uy\"ukkili\c c et al. \cite{5,6}. Relevant applications to the
blackbody problem can be found in \cite{7,8}. Preliminarily, we
define the $q$-deformed exponential and logarithm \cite{9}:
\begin{eqnarray}
\exp_q(x)={1\over[1-(q-1)x]^{\frac{1}{q-1}}} \ ,\hspace{2cm}
\ln_q(x)={1-(1/x)^{q-1} \over q-1} \ .
\end{eqnarray}
The deformed quantum distribution function is given by
\begin{eqnarray}
f={1\over1/\exp_{q}(-{\cal E})-\lambda} \ ,
\end{eqnarray}
where the parameter $q$ is greater or equal to 1 (in the last
case the standard statistics are recovered). Since
$F(f,\lambda)=\exp\{\ln_q[1/(1/f+\lambda )]\},$ from Eq.
(\ref{6}) we get
\begin{eqnarray}
\hat\varphi=\exp(\ln_q f) \ ,\hspace{2cm} \hat\psi=
\exp\left(\ln_qf-\ln_q{f\over 1+\lambda f}\right) \ .
\end{eqnarray}
We have $\lim_{f\to 0}\hat\varphi=0$ and $\lim_{f\to 0}
\hat\psi=1$ for $q<2$, otherwise it either vanishes (for
$\lambda<0$) or it diverges (for $\lambda>0$),
\begin{eqnarray}
\hat\psi(1,\,\lambda)=\exp\left(-\ln_q{1\over 1+\lambda}\right) \ ,
\end{eqnarray}
and
\begin{eqnarray}
\lim_{f\rightarrow\infty}\hat\psi(f,\,\lambda)=\exp\left(\frac{\lambda^{q-1}}{q-1}\right)
\ , \ \ \ {\rm for} \ \ \lambda>0 \ .
\end{eqnarray}
For fermions we observe that $\psi(1)=0$ is not fulfilled.
Moreover, in the case of bosons the condition
$\Psi(+\infty)=+\infty$ does not hold.\\
Figures 1a and 1b describe the couple of function $\varphi$ and
$\psi$ versus both $n$ and $q$. Figures 2a and 2b describe the
couple of functions $\Phi$ and $\Psi$ versus both $N$ and $q$.
Figure 1b shows clearly the violation of condition $\psi(1)=0$.
Both $\varphi$ and $\Phi$ start with a vanishing derivative,
while both $\psi$ and $\Psi$ start with an infinite derivative.\\
\begin{figure}[t]
\centerline{
\includegraphics[width=.8\columnwidth,angle=-90]{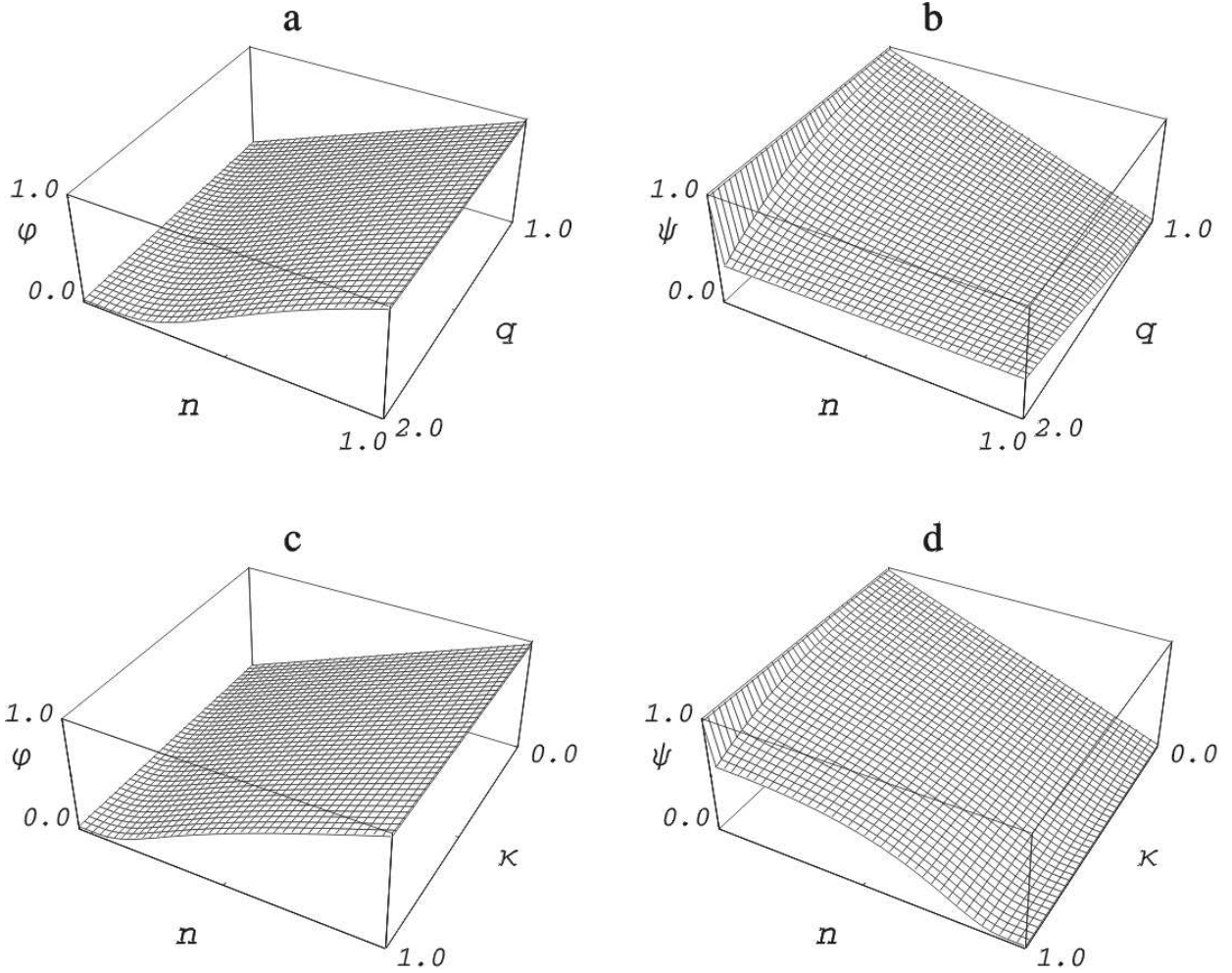}}
\caption{a) and b): plot of $\varphi$ and $\psi$ vs $n$ and $q$
for the model of B\"uy\"ukkili\c c et al; c) and d): plot of
$\varphi$ and $\psi$ vs $n$ and $\kappa$ for the model of
Kaniadakis.} \label{fig1}
\end{figure}
{\bf 2}) Recently Kaniadakis \cite{10,11} has proposed a new
classical distribution function, together with its quantum
counterparts. An application to Bose-Einstein condensation can be
found in \cite{12}. Preliminarily, we define the new
$\kappa$-deformed exponential and logarithm:
\begin{eqnarray}
\exp_{\{\kappa\}}(x)=\exp\left[{1\over\kappa}\sinh^{-1}(\kappa\,x)\right]
 \ ,\hspace{1.5cm} \ln_{\{\kappa\}}(x)={1\over\kappa}\,\sinh\left(\kappa\,\ln x\right) \ .
\end{eqnarray}
Observe the remarkable property $\exp_{\{\kappa\}}(-x)=1/\exp_{\{\kappa\}}(x)$.
The deformed quantum distribution function is given by
\begin{eqnarray}
f={1\over\exp_{\{\kappa\}}({\cal E})-\lambda} \ ,
\end{eqnarray}
where the parameter $\kappa$ is taken here greater or equal to 0
(in the last case the standard statistics are recovered). Since
$F(f,\lambda)=\exp[-\ln_{\{\kappa\}}(1/f+\lambda)]$, from Eq.
(\ref{6}) we obtain
\begin{eqnarray}
\hat\varphi=\exp(\ln_{\{\kappa\}} f) \ ,\hspace{2cm} \hat
\psi=\exp\left(\ln_{\{\kappa\}} f+\ln_{\{\kappa\}}{1+\lambda f\over f}\right) \ .
\end{eqnarray}
\begin{figure}[b]
\centerline{
\includegraphics[width=.8\columnwidth,angle=-90]{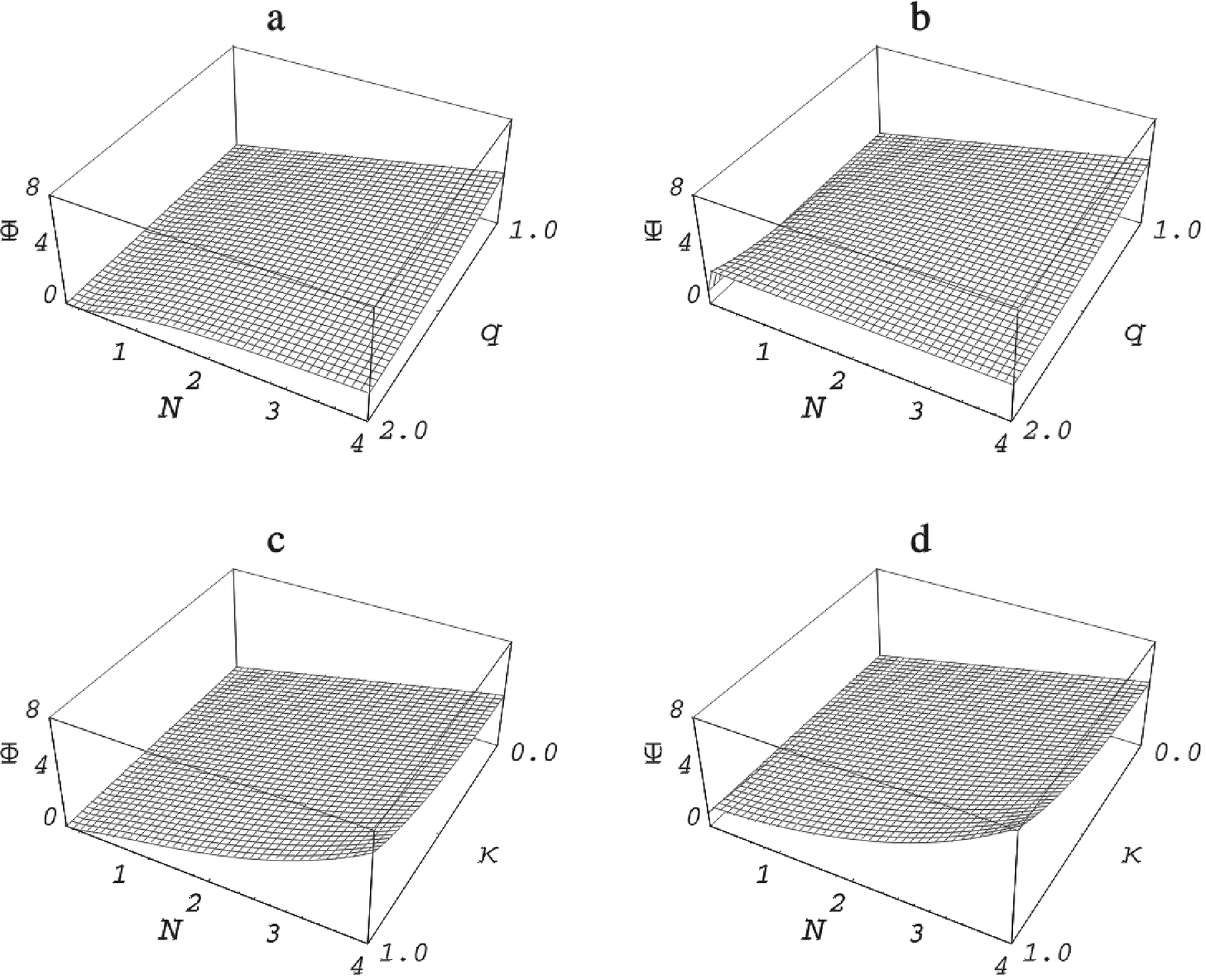}}
\caption{a) and b): plot of $\Phi$ and $\Psi$ vs $N$ and $q$ for
the model of B\"uy\"ukkili\c c et al; c) and d): plot of $\Phi$
and $\Psi$ vs $N$ and $\kappa$ for the model of Kaniadakis.}
\label{fig2}
\end{figure}
We have $\lim_{f\to 0}\hat\varphi=0$ and $\lim_{f\to 0}
\hat\psi=1$ for $\kappa<1$, otherwise, it either vanishes (for
$\lambda<0$) or it diverges (for $\lambda>0$), and
\begin{eqnarray}
\hat\psi(1,\,\lambda)=\exp[\ln_{\{\kappa\}}(1+\lambda)] \ .
\end{eqnarray}
For electrons, the last result agrees with the condition
$\psi(1)=0$. In the case of bosons, the condition for $\Phi$ when
$x\to\infty$ is fulfilled as well. Figure 1c and 1d describe the
couple of functions $\varphi$ and $\psi$ versus both $n$ and $\kappa$.
Figures 2c and 2d describe the couple of functions $\Phi$ and
$\Psi$ versus both $N$ and $\kappa$. Figure 1d shows clearly the
fulfillment of conditions $\psi(1)=0$. Both $\varphi$ and $\Phi$ start
with a vanishing derivatives, while both $\psi$ and $\Psi$ start
with an infinite derivative.

\section{Electrons in a phonon background: equilibrium and
stability}

A usual approximation to the BBP equations consists in
considering the relaxation due to p--p interactions to be much
quicker than the one due to e--p interactions. This assumption
amounts to fix, in the kinetic equation for electrons, the
distribution of phonons as the equilibrium function (\ref{1}) at
a given temperature $T$.\\ The kinetic equation for electrons
reads now
\begin{eqnarray}
\frac{\partial\,n_{\bf p}}{\partial\,
t}+{\bfm v}\cdot\frac{\partial\,n_{\bf p}}{\partial\,{\bfm x}}
-e\,{\bfm E}\cdot\frac{\partial\,n_{\bf p}}{\partial\,{\bfm
p}}=\left(\frac{\partial\,n_{\bf p}}{\partial\,t}\right)^\diamond_{ep} \
,\label{53}
\end{eqnarray}
where $\diamond$ means that $N_g^\ast$ has been substituted for
$N_g$. In order to study equilibrium and its stability, it is
useful to introduce the following functional:
\begin{eqnarray}
{\cal C}=2\,\int\left(\frac{\partial n_{\bf p}}{\partial
t}\right)_{ep}^\diamond\ln\left[\frac{\varphi(n_{\bf p})}{
\psi(n_{\bf p})}\,\exp\left(\frac{\epsilon_{\bf
p}}{T}\right)\right]\,d\,{\bfm p} \ ,\label{54}
\end{eqnarray}
which can be rewritten as follows:
\begin{eqnarray}
\nonumber {\cal C}&=&2\,\sum_g\int\int w_{ep}({\bfm
p}^\prime,{\bfm k}\rightarrow{\bfm p})\,\delta(\epsilon_{\bf
p}-\epsilon_{\bf p^\prime}-\omega_g)\,\ln
\left(\frac{\varphi^\prime\,\psi\,\Phi_g^\ast}{
\psi^\prime\,\varphi\,\Psi_g^\ast}\right)\\
&&\times[\Phi(N_g^\ast)\,\varphi(n_{{\bf p}^\prime})\,\psi(n_{\bf
p}) -\varphi(n_{\bf p})\,\psi(n_{{\bf
p}^\prime})\,\Psi(N_g^\ast)]\,\frac{d{\bfm k}\,d{\bfm
p}}{8\,\pi^3} \ \le\ 0 \ ,\label{55}
\end{eqnarray}
since $(A-B)\ln(B/A)\le0$.\\
In the space homogeneous and forceless case, equilibrium
solutions are given by
\begin{eqnarray}
\left(\frac{\partial\,n_{\bf p}}{\partial\,t}\right)_{ep}^\diamond=0 \
.\label{56}
\end{eqnarray}

\noindent{\bf Proposition 1.}

{\it Eq. (\ref{56}) is equivalent to}
\begin{eqnarray}
\Phi(N_g^\ast)\,\varphi(n_{{\bf p}^\prime})\,\psi(n_{\bf p})
=\varphi(n_{\bf p})\,\psi(n_{{\bf p}^\prime})\,\Psi(N_g^\ast) \
,\hspace{1cm} \forall \ {\bfm p} \ ,\, {\bfm k} \ .\label{57}
\end{eqnarray}

\noindent{\bf Proof.} In fact, we observe that Eq. (\ref{57}) $\
\Longrightarrow$ Eq. (\ref{56}) [see Eq. (\ref{15})]. On the other
hand, from Eq. (\ref{54}) it follows that Eq. (\ref{56})
$\Longrightarrow \ {\cal C}=0$. Tanks to Eq. (\ref{55}), we
obtain ${\cal C}=0\ \Longrightarrow$ Eq. (\ref{57}) $\bullet$

Eq. (\ref{57}), by taking into account Eq. (\ref{1}) and
$\epsilon_{\bf p}=\epsilon_{\bf p^\prime}+\omega_g$, shows that,
at equilibrium, $\ln[\varphi(n_{\bf p})/\psi(n_{\bf
p})]+\epsilon_{\bf p}/T$ is a collisional invariant for
electrons, that means $n_{\bf p}^\ast$ is given by Eq. (\ref{2}).
In order to study the stability of such an equilibrium solution,
we introduce the following functional
\begin{eqnarray}
L=H_e+{2\over T}\,\int\epsilon_{\bf p}\,n_{\bf p}\,d{\bfm p} \ ,
\end{eqnarray}
where $H_e=\int{\cal H}_e(n_{\bf p})\,d{\bfm p}$ and
$\partial{\cal H}_e/\partial n_{\bf p}=2\,\ln(\varphi/\psi)$
(observe that ${\cal H}_e$ is a convex function of $n_{\bf p}$,
since $\varphi/\psi$ has been assumed to be monotonically increasing).\\
We can prove an H theorem for the
present problem:

\noindent{\bf Proposition 2.}

{\it $L$ is a Lyapounov functional for the present problem.}

\noindent{\bf Proof.} By utilizing the definition of $L$ together
with Eq. (\ref{53}), one obtains $\dot L={\cal C}\ \le\ 0$, where
the dot means the time derivative. Moreover, since $(\partial{\cal
H}_e/\partial n_{\bf p})^\ast=2\,(\mu-\epsilon_{\bf p})/T$ by
taking into account electron conservation, we have
\begin{eqnarray}
\int\left(\frac{\partial{\cal H}_e}{\partial n_{\bf
p}}\right)^\ast(n_{\bf p}-n_{\bf p}^\ast)\,d{\bfm p}= -{2\over
T}\,\int\epsilon_{\bfm p}\,(n_{\bf p}-n_{\bf p}^\ast)\,d{\bfm p}
\ ,
\end{eqnarray}
so that we can write
\begin{eqnarray}
L-L^\ast=\int\left\{{\cal H}_e-\left[{\cal H}_e^\ast
+\left(\frac{\partial{\cal H}_e}{\partial n_{\bf p}}\right)^\ast
\,(n_{\bf p}-n_{\bf p}^\ast)\right]\right\}\,d{\bfm p}\ \ge0 \ ,
\end{eqnarray}
due to the convexity of ${\cal H}_e \ \bullet$

Let us now interpret this result on a physical ground. First, we
define the concentration N, the energy density $E_e$, and the
entropy density $S_e$ of the electron gas as follows:
\begin{eqnarray}
{\rm N}={1\over 8\,\pi^3}\int 2\,n_{\bf p}\,d{\bfm p} \ ,\hspace{1cm}
E_e={1\over 8\,\pi^3} \int2\,n_{\bf p}\epsilon_{\bf p}\,d{\bfm p} \
,\hspace{1cm} S_e=-{H_e\over 8\,\pi^3} \ ,
\end{eqnarray}
where the factor 2 inside these integrals accounts for
degeneracy. Observe that this definition of $S_e$ is consistent
with the definitions of $T$ and $\mu$, which we have already
given. In fact, at equilibrium, the following thermodynamical
relationships \cite{Reif} are recovered:
\begin{eqnarray}
\left(\frac{\partial S_e}{\partial {\rm N}}\right)_{E_e}=\
-\frac{\mu}{T} \ ,\hspace{2cm} \left(\frac{\partial S_e}{\partial
E_e}\right)_{\rm N}=\ {1\over T} \ .
\end{eqnarray}
Now, from the definition of $L$, it is easy to realize that the
meaning of $\dot L\le0$ is nothing but the Clausius inequality
$\dot S_e\ \ge\ \dot E_e/T.$

\newpage  \noindent{\bf ACKNOWLEDGMENT}

This work has been supported by the Fonds zur F\"orderung der
wissenschaftlichen Forschung, Vienna, under contract No.
P14669-TPH.

\vfill\eject
\end{document}